      [1999/12/01 v1.4c Il Nuovo Cimento]
\documentclass[final]{cimento}

\usepackage{graphicx}  
\title{The relative concentration of visible and dark matter in clusters of galaxies}
\shorttitle{Dark matter in clusters of galaxies}
\author{C.~De Boni\from{ins:x} \from{ins:y}\thanks{Mail: cristiano.deboni@unibo.it}
        \atque
G.~Bertin\from{ins:x}
                                              }
\instlist{\inst{ins:x} Universit\`a degli Studi di Milano, Dipartimento di Fisica, via Celoria 16, I-20133 Milano, Italy
  \inst{ins:y} Universit\`a di Bologna, Dipartimento di Astronomia, via Ranzani 1, I-40127 Bologna, Italy
               }
\PACSes{\PACSit{95.35.+d}{Dark matter (stellar, interstellar, galactic, and cosmological)}
\PACSit{98.65.Cw}{Galaxy clusters}
\PACSit{98.65.Hb}{Intracluster matter, cooling flows}}
\begin{document}

\maketitle

\begin{abstract}
In general, the best evidence for the presence of dark matter is
obtained when an additional mass component with distribution
different from that of the visible matter is needed to explain the
dynamical data. Here we wish to study the relative distribution of
visible and dark matter in clusters of galaxies, in comparison
with the distribution observed on the galactic scale; in
particular, we wish to check whether dark matter is more
concentrated than visible matter without using assumptions or
other constraints deriving from cosmology. We also intend to check
whether the result depends significantly on the dynamical state of
the cluster under investigation. We consider two clusters (A496
and Coma) that are representative of the two classes of cool-core
and non-cool-core clusters. We first refer to a two-component
dynamical model that ignores the contribution from the galaxy
density distribution and study the condition of hydrostatic
equilibrium for the hot intracluster medium (ICM) under the
assumption of spherical symmetry, in the presence of dark matter.
We model the ICM density distribution in terms of a standard
$\beta$-model with $\beta=2/3$, i.e. with a distribution similar
to that of a regular isothermal sphere (RIS), and fit the observed
X-ray brightness profiles. With the explicit purpose of ignoring
cosmological arguments, we na\"ively assume that dark matter, if
present, has an analogous density distribution, with the freedom
of two different density and length scales. The relative
distribution of visible and dark matter is then derived by fitting
the temperature data for the ICM under conditions of hydrostatic
equilibrium. For both clusters, we find that dark matter is more
concentrated with respect to visible matter. We then test whether
the conclusion changes significantly when dark matter is taken to
be distributed according to cosmologically favored density
profiles and when the contribution of the mass contained in
galaxies is taken into account. Although the qualitative conclusions remain unchanged, we find that the contribution of galaxies to the mass budget is more important than generally assumed. We also show that, without
resorting to additional information on the small scale, it is not
possible to tell whether a density cusp is present or absent in
these systems. In contrast with the case of individual galaxies,
on the large scale in clusters of galaxies dark matter is indeed
more concentrated than visible matter.
\end{abstract}

\section{Introduction} \label{one}

Dark matter is known to play a significant role on the scale of
galaxies and to dominate on the scale of clusters of galaxies.

On the galactic scale, for both spirals and ellipticals, the
contribution of dark matter to the total mass generally exceeds
50\%. Typically, the dark component is diffuse, i.e. it becomes
more important at large radii (e.g., see Bertin \cite{bertin00}).
This is the reason why in galaxies it is usually referred to as a
dark matter ``halo''.

On the larger scale of clusters of galaxies, it is generally
stated that dark matter represents about 85\% of the total mass
and that the visible matter is mostly in the form of a hot
intracluster medium (ICM) (for some estimates see, e.g., Mohr et
al. \cite{mohr99}; Ettori et al. \cite{ettori02}; Rosati et al.
\cite{rosati02}). Much work has been done to study how this
fraction changes with radius (e.g., David et al. \cite{david95};
White \& Fabian \cite{white95}; Ettori \& Fabian \cite{ettori99};
Vikhlinin et al. \cite{vikhlinin06}; Allen et al. \cite{allen08}).

In this paper, we wish to study the relative distribution of dark
and visible matter in clusters of galaxies in the most simple
dynamical framework. On purpose, we will ignore any suggestion or
constraint from the cosmological context and proceed by applying a
most na\"ive and simple parametric dynamical model to the hot ICM,
considered to be in hydrostatic equilibrium.

The work presented here is based on the use of X-ray data from
\emph{BeppoSAX}, which are well suited for the discussion of the
mass distribution on the large scale, at least out to a scale
close to 1 Mpc. Clearly, for the discussion of the mass
distribution on the small scale of the inner sphere of radius
$\approx$ 100 kpc, other data (X-ray data from \emph{Chandra},
strong lensing, or stellar dynamical data for the bright central
galaxy often present) would provide the decisive pieces of
information but would also require a much more sophisticated
modeling (see Mahdavi et al. \cite{mahdavi07}). Most likely, in
order to make a significant statement beyond the scale of 1 Mpc,
different diagnostics, such as that offered by the
Sunyaev-Zel'dovich effect, will be required.

The paper is organized as follows. In Sect. \ref{methods} the
simplest dynamical model (in which the contribution of the mass
associated with galaxies is ignored) is introduced. The results
are presented in Sect. \ref{three} and discussed in Sect.
\ref{four}. The contribution of the mass associated with galaxies
is discussed in Sect. \ref{five}. In Sect. \ref{six} the
conclusions are drawn.

To quantify the scales of the density distributions in the two
clusters considered, except where explicitly noted, we refer to
the following value of the Hubble constant: $H_{0} = 50 \rm{\ km \
s^{-1} \ Mpc^{-1}}$. This is equivalent to setting $h = 0.5$,
where $h$ denotes the Hubble constant in units of $100 \rm{\ km \
s^{-1} \ Mpc^{-1}}$. This choice is rather common within the X-ray
astrophysics community. In Sect. \ref{four} we comment on the
consequences of adopting the currently favored value $h = 0.7$.

\section{The simplest dynamical model} \label{methods}

Even if the morphology of galaxy clusters is quite complex, we
adopt spherical symmetry for the ICM, on the scales considered. We
assume hydrostatic equilibrium and a local ideal equation of
state, i.e. we ignore the possible presence of turbulent pressure.
We study a model with only two components, i.e. the intracluster
medium and the dark matter. For simplicity, the contribution of
galaxies is neglected. Strictly speaking, what we call dark matter
in our model, being the difference between total and ICM matter,
also includes the contribution of some visible mass. In practice,
given our focus on the large-scale distributions, our simplified
notation is appropriate. We will return on this issue in Sect.
\ref{five}, where we discuss the mass contribution associated with
the galaxies.

Starting from these hypotheses, we now proceed to describe how a
dynamical model can be constructed so that the relative mass
distribution in clusters of galaxies can be derived from the X-ray
data. First of all, we extract the ICM radial density profiles by
fitting the electron density distribution (associated with the
X-ray brightness profiles) with a $\beta$-model (Cavaliere \&
Fusco-Femiano \cite{cavaliere76}).

We take (e.g., see Voit \cite{voit05} and references therein)
$\beta=2/3$ and express the radial coordinate in dimensionless
form  as $x=r/r_{X}$ so that our fitting formula can be written as

\begin{equation}
\rho_{X}(x)= {\rho_{X}}^{(0)} \left(1+x^2\right)^{-1} \ ;
\label{ICM}
\end{equation}

\noindent here ${\rho_{X}}^{(0)}$ and $r_{X}$ are the density and
the length scales. Note that this is often used as a simple
representation of the density profile of a self-gravitating
regular isothermal sphere (RIS).

We then consider the hydrostatic equilibrium equation

\begin{equation}
\frac{k}{\mu m _{p}} \frac{1}{\rho _{X}(r)} \frac{d}{dr} \left[ \rho _{X}(r) T _{X}(r) \right]=- \frac{GM_{tot}(r)}{r^{2}} \ ,
\end{equation}

\noindent where $T_{X}(r)$ is the value of the ICM temperature at
a given distance $r$ from the center of the cluster and
$M_{tot}(r)$ is the total mass of the system at radius $r$

\begin{equation}
M_{tot}(r)=4\pi \int_{0}^{r} \left[\rho_{X}(r') + \rho_{DM}(r') \right]  r'^{2}dr'=M_{X}(r)+M_{DM}(r) \ .
\label{Mtot}
\end{equation}

\noindent Here $\rho_{DM}(r)$ is identified with the dark matter
density profile. We may then introduce the quantity

\begin{equation}
f_{gas}(r) = \frac{M_{X} (r)}{M_{tot} (r)} \ ,
\label{f(r)}
\end{equation}

\noindent which represents the gas fraction and is an increasing
function of radius if dark matter is more concentrated with
respect to visible matter.

To test whether dark matter is more or less concentrated with
respect to visible matter, we adopt the simplest form for the dark
matter density profile, i.e. the approximate representation of a
regular isothermal sphere (RIS),

\begin{equation}
\rho_{DM}(x)= {\rho_{DM}}^{(0)} \left[1+\left(\frac{x}{\lambda}\right)^2\right]^{-1} \ ,
\label{RIS}
\end{equation}

\noindent where we have introduced a ``concentration parameter''
$\lambda = r_{DM}/ r_{X}$. If the fitting procedure will yield
values of $\lambda$ smaller or larger than unity, i.e. values of
$r_{DM}$ smaller or larger than $r_{X}$, we will say that dark
matter is more or less concentrated, respectively.

For comparison, we will also refer to a NFW profile (Navarro et
al. \cite{navarro97}), which we express as

\begin{equation}
\rho_{NFW}(x)= \frac{{\rho_{NFW}}^{(0)}}{(x/\lambda_{c})(1+x/\lambda_{c})^2} \ .
\label{NFW}
\end{equation}

\noindent Note that  $\lambda_{c}$ is not a direct measure of the
dark matter concentration, because the $\beta$-model for the ICM
and the NFW profile for the dark matter have different analytical
forms. Since the models considered (Eq. (\ref{ICM}), Eq.
(\ref{RIS}), and Eq. (\ref{NFW})) generate mass profiles that
diverge at large radii, we introduce a cutoff radius $r_{out} =
ar_{X}$ (to be interpreted as the outermost radius for which
observational X-ray temperature and brightness data are
available). We thus introduce the parameter $\textsf{f}$ as the
visible-to-dark mass content ratio over the whole cluster volume
defined by $r_{out}$:

\begin{equation}
\int_{0}^{ar_{X}} \rho _{X}(r)   r^{2}dr = \textsf{f} \int_{0}^{ar_{X}} \rho _{DM} (r) r^{2}dr \ .
\end{equation}

\noindent In this way the dimensional scales of the dark matter
density profile (either Eq. (\ref{RIS}) or Eq. (\ref{NFW})) can be
expressed through the dimensionless parameters $\lambda$ and
$\textsf{f}$. In terms of the function $f_{gas}$ introduced in Eq.
(\ref{f(r)}), we thus have $f_{gas}(r_{out})=
\textsf{f}/(\textsf{f}+1)$.

The hydrostatic equilibrium equation is then used to derive a
theoretical evaluation of the ICM temperature at any given radius,
$T_{X}(x,\lambda,\textsf{f})$. This temperature value can be
compared to the observed one, $T_{X}^{obs}(x)$. By minimizing

\begin{equation}
\chi^2 = \sum_{i=1}^{n} \left[ \frac{T_{X}^{obs}(x_{i}) - T_{X}(x_{i},\lambda,\textsf{f})}{\sigma_{i}^{T}} \right]^{2} \
\label{chi2temperature}
\end{equation}

\noindent where $x_{i}$ are the locations where the temperature
data are available, $x_{n}=a$, and $\sigma_{i}^{T}$ is the error
in $T_{X}^{obs}(x_{i})$, the best-fit values of the parameters
$\lambda$ and $\textsf{f}$ are found. Note that, in this approach,
we are making no assumptions on the possible existence of a global
equation of state (such as the polytropic equation $p \propto
\rho^{\gamma}$), which might dictate a given profile $T_{X}$ in
correspondence of a given density profile $\rho_{X}$ of the ICM.
There is growing consensus that, in the outer regions, the ICM has
a declining temperature profile (Markevitch et al.
\cite{markevitch98}; De Grandi \& Molendi \cite{degrandi02};
Vikhlinin et al. \cite{vikhlinin05}, \cite{vikhlinin06}; Pratt et
al. \cite{pratt07}), but here we need not enter the issue of
whether this profile has universal character and can be traced to
a general equation of state.

\subsection{The sample and the data} \label{sample}

We have decided to test this method by focusing on two clusters
(A496 and A1656, i.e. Coma) which are representative of two
classes: cool-core (or relaxed) clusters and non-cool-core (or
non-relaxed) clusters. Both clusters belong to the nearby
universe, at redshift $z=0.0320$ and $0.0232$, respectively.

For the ICM data, we take those of Ettori et al. \cite{ettori02},
where the methods for collecting the data and their deprojection
are explained in detail (see also De Grandi \& Molendi
\cite{degrandi01}, \cite{degrandi02}).

Even if spherical symmetry and hydrostatic equilibrium may be
inadequate when dealing with accurate models of specific cases,
especially in the case of non-relaxed clusters (for a relatively
recent study of the dynamical state of Coma, see Neumann et al.
\cite{neumann03}), the two clusters considered in this paper are
sufficiently regular to be taken as reasonable test-cases for the
application of our simple dynamical model to the study of the
relative concentration of visible and dark matter on the large
scale.

\section{Results} \label{three}

\subsection{Electron density profiles} \label{electron}

For the electron density profile we follow Eq. (\ref{ICM}), with
electron density scale $n_{0}$ and obtain the values of the model
parameters (the central electron density scale $n_{0}$ and the
length scale $r_{X}$) by minimizing

\begin{equation}
{\chi}^2= \sum_{i=1}^{n} \left[ \frac{n_{e}^{obs} (r_{i}) - n_{e} (r_{i})}{\sigma_{i}^n} \right]^2 \ ,
\end{equation}

\noindent where $r_{1}$ is the radius of the innermost bin for
which we have a density measure and $n_{e}^{obs} (r_{i})$  is the
observed density value at $r_{i}$; $\sigma_{i}^n$ is the
uncertainity in $n_{e}^{obs}(r_{i})$.

In order to avoid modeling problems concerning the possible
presence of turbulent motions or the physics of individual
galaxies, we have decided to make a 100 kpc cut in the central
regions. This leads to the exclusion of the innermost data point.
Therefore, the central density $n_{0}$ is the extrapolation to the
center of this fit and should not be taken literally as an
estimate of the central electron density.

In Table \ref{electrontable} we report the best-fit values for
each cluster; these values are the ones that are used in Subsect.
\ref{temperature} to derive the dark matter parameters.

\begin{table}[!h]
\caption{Best-fit values for the parameters of the electron
density profile (Eq. (\ref{ICM})) for A496 and Coma ($h=0.5$).} 
\centering
\begin{tabular}{c c c c c c}
\hline
\\
Cluster & {$n_{0}$} & {${\rho_{X}}^{(0)}$} & $r_{X}$ & ${\tilde{\chi}}^2$ & (d.o.f.) \\
& {($10^{-3} \ {\rm{cm}}^{-3}$)} & {$(10^{14} \ {\rm{M_{\odot}}} \ {\rm{Mpc}}^{-3})$} & (Mpc) & \multicolumn{2}{c}{}  \\
\\
\hline
\\
A496 & 4.$910_{-0.060}^{+0.047}$ &  1.411 & $0.182_{-0.002}^{+0.002}$ & 4.9 & (3) \\
\\
Coma & 2.$987_{-0.020}^{+0.018}$ &  0.8584 & $0.410_{-0.002}^{+0.002}$ & 40.3 & (3) \\
\\
\hline
\end{tabular}
\label{electrontable}
\end{table}

In the third column of Table \ref{electrontable} we record the
values of the central mass density of the ICM, converted from the
electron density with the relation $\rho_{X}=\mu_{e} n_{e} m_{p}$,
with $m_{p}$ the proton mass and $\mu_{e}=1.1696$.

In both cases, but especially for Coma (see also our final comment
in Subsect. \ref{sample}), the values of the reduced ${{\chi}}^2$,
denoted by ${\tilde{\chi}}^2$, are quite high: this is due to the
fact that the values of $\sigma^n$ are small, since the measures
of the electron density are very precise. We will use the model
and the inferred parameters as a reasonable representation of the
electron density in the region outside the central 100 kpc.

\subsection{Temperature profiles} \label{temperature}

Based on the input parameters obtained in Subsect. \ref{electron},
we have proceeded to apply the method described in Sect.
{\ref{methods}. The temperature profiles are illustrated in Fig.
{\ref{temperatureprofiles}.

In Table \ref{bestfitA496} and in Table \ref{bestfitA1656} we
report the best-fit values for the dark matter parameters
$\lambda$ and $\textsf{f}$, obtained by minimizing Eq.
(\ref{chi2temperature}) using either Eq. (\ref{RIS}) (which is
listed as RIS) or Eq. (\ref{NFW}). We recall that for the RIS
model, the parameter $\lambda$ provides a direct indication of the
relative concentration of the dark matter.

\begin{table}[!h]
\caption{Best-fit values for the parameters of the dark matter distribution for A496 ($h=0.5$).}
\centering
\begin{tabular}{c c c c c}
\hline
\\
Model & $\lambda$ & \textsf{f} & $\tilde{\chi}^2$ & (d.o.f.) \\
\\
\hline
\\
RIS & $0.50_{-0.02}^{+0.02}$ & $0.135_{-0.001}^{+0.002}$ & 0.8459 & (2) \\
\\
NFW & $3.15_{-0.11}^{+0.11}$ & $0.132_{-0.002}^{+0.001}$ & 0.7823 & (2) \\
\\
\hline
\end{tabular}
\label{bestfitA496}
\end{table}

\begin{table}[!h]
\caption{Best-fit values for the parameters of the dark matter distribution for Coma ($h=0.5$).}
\centering
\begin{tabular}{c c c c c}
\hline
\\
Model & $\lambda$ & \textsf{f} & $\tilde{\chi}^2$ & (d.o.f.) \\
\\
\hline
\\
RIS & $0.30_{-0.01}^{+0.01}$ & $0.161_{-0.001}^{+0.003}$ & 0.6920 & (3) \\
\\
NFW & $3.26_{-0.14}^{+0.19}$ & $0.143_{-0.002}^{+0.002}$ & 0.5270 & (3) \\
\\
\hline
\end{tabular}
\label{bestfitA1656}
\end{table}

Here the values of $\tilde{\chi}^2$ are all below unity. Note that
the values of $\textsf{f}$ are quite similar for the RIS model and
the NFW profile. From inspection of the RIS model results, we see
that for both clusters the value of $\lambda$ is less than one, a
clear indication that dark matter is more concentrated than
visible matter. For the NFW profile, our values of $\lambda$ are
consistent with those reported by Ettori \& Fabian
\cite{ettori99}.

\section{Discussion} \label{four}

In order to better appreciate the relative concentration of dark
and visible matter for the models identified in Tables
\ref{bestfitA496} and \ref{bestfitA1656}, we may refer to the
integrated mass profiles introduced in Sect. \ref{methods} (see
Eqs. (\ref{Mtot}), (\ref{f(r)})). The various profiles are
illustrated in Fig. \ref{massprofiles}. Tables \ref{A496masses}
and \ref{A1656masses} summarize some characteristics of the
best-fit models.

\begin{table}[!h]
\caption{Mass associated with the ICM ($M_{X}$), mass associated with the dark matter ($M_{DM}$), total mass ($M_{tot}$), and gas fraction ($f_{gas}$) evaluated at $r_{out}=743 \ \rm{kpc}$, for the best-fit RIS model and NFW profile for A496 ($h=0.5$).}
\centering
\begin{tabular}{c c c c c}
\hline
\\
Model & $M_{X}(r_{out})$ & $M_{DM}(r_{out})$ & $M_{tot}(r_{out})$ & $f_{gas}(r_{out})$ \\
& ($10^{14} \ {\rm{M}_{\odot}}$) & ($10^{14} \ {\rm{M}_{\odot}}$) & ($10^{14} \ {\rm{M}_{\odot}}$) & \\
\\
\hline
\\
RIS & 0.294 & 2.18 & 2.47 & 0.119 \\
\\
NFW & 0.294 & 2.24 & 2.53 & 0.116 \\
\\
\hline
\end{tabular}
\label{A496masses}
\end{table}

\begin{table}[!h]
\caption{Mass associated with the ICM ($M_{X}$), mass associated with the dark matter ($M_{DM}$), total mass ($M_{tot}$), and gas fraction ($f_{gas}$) evaluated at $r_{out}=702 \ \rm{kpc}$, for the best-fit RIS model and NFW profile for Coma ($h=0.5$).}
\centering
\begin{tabular}{c c c c c}
\hline
\\
Model & $M_{X}(r_{out})$ & $M_{DM}(r_{out})$ & $M_{tot}(r_{out})$ & $f_{gas}(r_{out})$ \\
& ($10^{14} \ {\rm{M}_{\odot}}$) & ($10^{14} \ {\rm{M}_{\odot}}$) & ($10^{14} \ {\rm{M}_{\odot}}$) & \\
\\
\hline
\\
RIS & 0.498 & 3.09 & 3.58 & 0.139 \\
\\
NFW & 0.498 & 3.49 & 3.98 & 0.125 \\
\\
\hline
\end{tabular}
\label{A1656masses}
\end{table}

Figure \ref{massprofiles} shows that, in each cluster, the gas
fraction increases with the distance from the center, i.e. the
dark matter is more concentrated than the visible matter. Note
that this statement holds for two very different models of the
density distribution of dark matter, which turn out to exhibit
only modest quantitative differences in relation to the issues
addressed in this paper.

As noted at the end of Sect. \ref{one}, in this paper the analysis
has been carried out under the assumption of a rather low Hubble
constant, with $h=0.5$. It is easy to convert the relevant
quantitative estimates to those appropriate for a different value
of $h$ by recalling (e.g., see \cite{mohr99}) that lengths (such
as $r_{X}$ of Table \ref{electrontable}) scale as $h^{-1}$, gas
densities ($n_{0}$ and ${\rho_{X}}^{(0)}$ of Table
\ref{electrontable}) as $h^{1/2}$, the integrated ICM mass (see
$M_{X}$ in Tables \ref{A496masses} and \ref{A1656masses}) as
$h^{-5/2}$, the total binding mass (see $M_{tot}$ in Tables
\ref{A496masses} and \ref{A1656masses}) as $h^{-1}$, and finally
the gas fraction (see $f_{gas}$ in Tables \ref{A496masses} and
\ref{A1656masses}) as $h^{-3/2}$. Of course $\lambda$ is
independent of $h$.

For example, for $h=0.7$ we would have for Coma $r_{out}= \rm{501
\ kpc}$ (instead of 702 kpc) and a cut in the central regions
$r_{cut}= \rm{71 \ kpc}$ (instead of 100 kpc); the length scales
for the best-fit RIS model and NFW profile along with the
integrated masses associated with them are reported in Table
\ref{A1656masses(h=0.7)}.

\begin{table}[!h]
\caption{Length scales for the ICM ($r_{X}$) and the dark matter ($r_{DM}$), mass associated with the ICM ($M_{X}$), mass associated with the dark matter ($M_{DM}$), total mass ($M_{tot}$), and gas fraction ($f_{gas}$) evaluated at $r_{out}=501 \ \rm{kpc}$, for the best-fit RIS model and NFW profile for Coma, for a Hubble constant corresponding to $h=0.7$.}
\centering
\begin{tabular}{c c c c c c c}
\hline
\\
Model & $r_{X}$ & $r_{DM}$ & $M_{X}(r_{out})$ & $M_{DM}(r_{out})$ & $M_{tot}(r_{out})$ & $f_{gas}(r_{out})$ \\
& (Mpc) & (Mpc) & ($10^{14} \ {\rm{M}_{\odot}}$) & ($10^{14} \ {\rm{M}_{\odot}}$) & ($10^{14} \ {\rm{M}_{\odot}}$) & \\
\\
\hline
\\
RIS & 0.293 & 0.088 & 0.215 & 2.35 & 2.56 & 0.084 \\
\\
NFW & 0.293 & 0.955 & 0.215 & 2.63 & 2.84 & 0.076 \\
\\
\hline
\end{tabular}
\label{A1656masses(h=0.7)}
\end{table}

\section{The role of the galaxies} \label{five}

The relatively low values of $f_{gas}$ recorded in Table
\ref{A1656masses(h=0.7)} suggest that we should check the role of
the galaxies in the mass budget of a cluster. In fact, so far we
have neglected the explicit contribution of galaxies or, rather,
absorbed their contribution into the dark matter component of the
mass distribution.

We focus on the case of Coma. In their study based on $h=0.7$,
{\L}okas \& Mamon \cite{lokas03} fit the three-dimensional
luminosity density profile of the galaxies in Coma with a NFW
profile and then obtain the mass distribution of the galaxies by
multiplying by a mean mass-to-light ratio (blue band) appropriate
for the morphological population of the cluster,
$\Upsilon_{G}=6.43 \ \rm{M_{\odot}/L_{\odot}}$. They find a length
scale for the galaxy distribution $r_{s}=0.411 \ \rm{Mpc}$. For a direct comparison with the simple modeling considered in
our paper, we have referred to a RIS profile

\begin{equation}
\rho_{G}(x)= {\rho_{G}}^{(0)} \left[1+\left(\frac{x}{\lambda_{G}}\right)^2\right]^{-1} \ ,
\label{galaxiesRIS}
\end{equation}

\noindent where $\lambda_{G}= r_{G}/r_{X}$, being $r_{G}$ the
length scale of the galaxies. We have found that for $r_{G}=0.071 \ \rm{Mpc}$,
corresponding to  $\lambda_{G}= 0.24$, and actually equal to
$r_{cut}$, such profile gives a similarly adequate representation of the galaxy
density distribution in the cluster. These latter values should be
compared with the dark matter distribution length scale and
concentration parameter we found for the RIS profile (see Tables
\ref{A1656masses(h=0.7)} and \ref{bestfitA1656}), $r_{DM}= 0.088 \
\rm{Mpc}$ and $\lambda=0.30$ respectively. This shows that, in the
Coma cluster, galaxies are somewhat more concentrated with respect
to dark matter and confirms the argument that led to the choice of
the cut in the inner regions, as discussed in Sect. 3. On the
other hand, we see that the difference between $\lambda_{G}= 0.24$
and $\lambda=0.30$ is rather small, so that, {\it a posteriori},
our choice of combining together, into a unique density profile
$\rho_{DM}(r)$, dark matter and the smaller contribution of matter
associated with galaxies was indeed very reasonable.

In order to quantify the effects of neglecting the galaxy
contribution, in Table \ref{A1656masses3c(h=0.7)} we report, for
$r_{cut}$ and $r_{out}$, the mass of the galaxies and the actual
mass of dark matter (i.e. the difference between the mass of what
we have called dark matter so far and the mass of the galactic
component), in addition to the mass of the ICM and the total mass,
together with the gas fraction and the galactic fraction

\begin{equation}
f_{G}(r)=\frac{M_{G}(r)}{M_{tot}(r)} \ ;
\end{equation}

\noindent here $M_{G}(r)$ is the cumulative mass obtained by integrating Eq. (\ref{galaxiesRIS}).

\begin{table}[!h]
\caption{Mass associated with the galaxies ($M_{G}$), mass
associated with the ICM ($M_{X}$), actual mass associated with the
dark matter ($M_{DM}^{actual}$), total mass ($M_{tot}$), galactic
fraction ($f_{G}$), and gas fraction ($f_{gas}$) evaluated at
$r_{cut}=71 \ \rm{kpc}$ and $r_{out}=501 \ \rm{kpc}$, for the
best-fit RIS model for Coma, for a Hubble constant corresponding
to $h=0.7$.} 
\centering
\begin{tabular}{c c c c c c c}
\hline
\\
Radius & $M_{G}(r)$ & $M_{X}(r)$ & $M_{DM}^{actual}(r)$ &
$M_{tot}(r)$ & $f_{G}(r)$ & $f_{gas}(r)$ \\ (Mpc) & ($10^{14} \
{\rm{M}_{\odot}}$) & ($10^{14} \ {\rm{M}_{\odot}}$) & ($10^{14} \
{\rm{M}_{\odot}}$) & ($10^{14} {\rm{M}_{\odot}}$) & & \\
\\
\hline
\\
0.071 & 0.00208 & 0.00147 & 0.0646 & 0.0682 & 0.030 & 0.022 \\
\\
0.501 & 0.0545 & 0.215 & 2.29 & 2.56 & 0.021 & 0.084 \\
\\
\hline
\end{tabular}
\label{A1656masses3c(h=0.7)}
\end{table}

We note from Table \ref{A1656masses3c(h=0.7)} that the
contribution by the galaxies to the total mass is at most of 3\%
at every radius outside $r_{cut}$; we also see (cf. Fig. \ref{Comaprofiles(h=0.7)}) that the fraction
of the sum $f_{G}(r) + f_{gas}(r)$ increases with radius,
confirming again that dark matter is concentrated with respect to
visible matter, even when we include the contribution of the mass
associated with galaxies.

Finally, we have tested the effect of increasing the value of the
mass-to-light ratio associated with galaxies, above the value
adopted by {\L}okas \& Mamon \cite{lokas03}, as would be
appropriate to do if galaxies were associated with increasingly
dominant dark matter halos. The combined value $f_{G}(r) +
f_{gas}(r)$ remains definitely a monotonic increasing function of
radius even when we double the galaxy mass-to-light ratio. Only at
$\Upsilon_{G} \approx 30 \ \rm{M_{\odot}/L_{\odot}}$ does the
trend start to reverse.

\section{Conclusions} \label{six}

In this work we have studied the relative distribution of the ICM
mass with respect to the total mass for two galaxy clusters. These
clusters (A496 and Coma) were chosen as representative of the
classes of cool-core and non-cool-core clusters respectively.

The main point of the paper is that some interesting conclusions
can be drawn by means of an extremely simple analysis. They are
the following:
\begin{enumerate}
\item We find that dark matter is concentrated with respect to visible matter. This result confirms previous investigations (see Markevitch et al. \cite{markevitch99}), but has been obtained here without imposing a global equation of state for the ICM or a cosmologically suggested density profile for the dark matter distribution. Therefore, our simple result contributes significant confidence in the robustness of this general conclusion.
\item On the large scale explored by the data used and the dynamical model adopted, the RIS model and the NFW profile are substantially equivalent and we cannot really tell whether a cusp of dark matter is present or absent. This conclusion basically confirms a natural expectation.
\item The two clusters studied here, although very different from the morphological point of view, exhibit a similar behaviour with respect to the issue addressed in this paper (see also Ettori et al. \cite{ettori02}).
\item For the currently favored value of the Hubble constant, $h=0.7$, the relative contribution
of galaxies to the mass budget requires a detailed check, because,
with respect to the case $h=0.5$, it is less obvious that on the
scale of $0.5~\rm{Mpc}$ the visible mass be dominated by the
intracluster medium. A quantitative test for the Coma cluster has
shown that indeed the contribution of galaxies is dominant only
in the innermost regions excluded by our cut to the data, inside
$0.1~\rm{Mpc}$. The conclusion that dark matter is more
concentrated than visible matter remains valid even when the
contribution of the galactic component to the visible mass is
considered explicitly.
\end{enumerate}

\acknowledgments We wish to thank Stefano Ettori for providing us
with the X-ray data at the basis of this investigation and for
several useful suggestions.

\newpage

\begin{figure}[!h]
\begin{tabular}{c c}
\includegraphics[width=65mm]{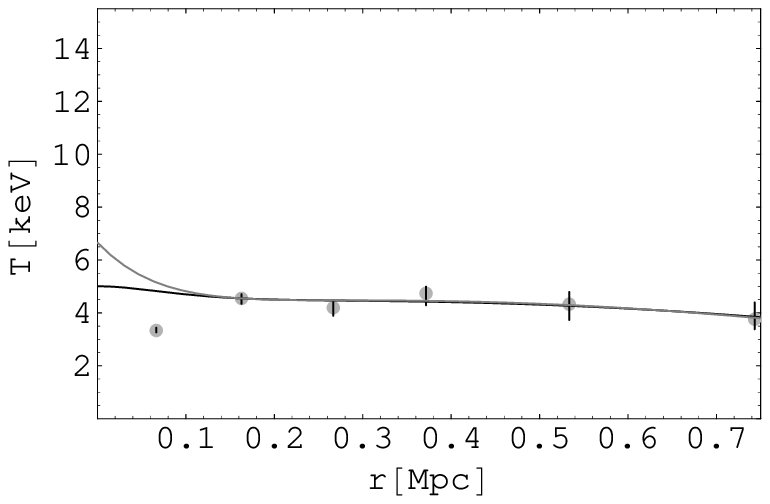} & \includegraphics[width=65mm]{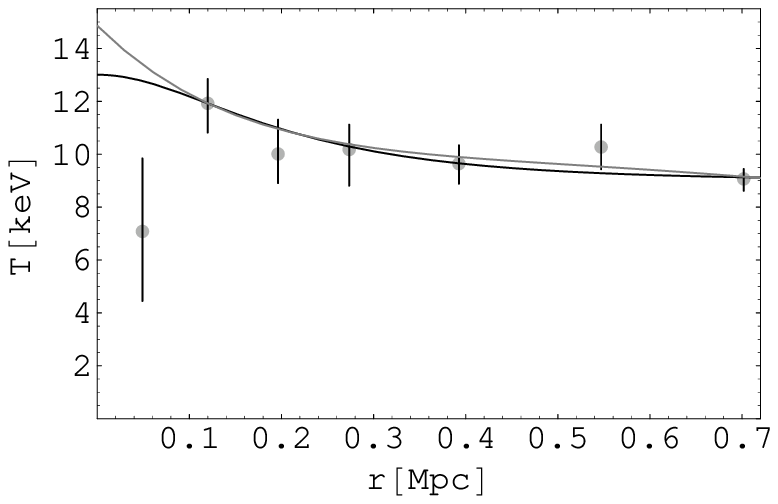}
\end{tabular}
\caption{Temperature profiles for A496 (\emph{left}) and Coma (\emph{right}) for the RIS model (\emph{thick line}) and the NFW profile (\emph{thin line}). The \emph{circles} are the data with the error bars; for both clusters, the innermost data point is inside the 100 kpc cut region, so it is excluded from the fit. The figure is based on $h=0.5$.}
\label{temperatureprofiles}
\end{figure}

\begin{figure}[!h]
\begin{tabular}{c c}
\includegraphics[width=65mm]{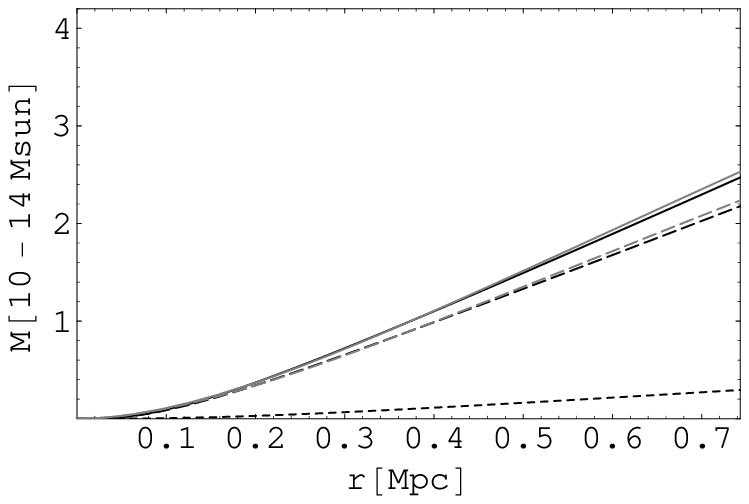} & \includegraphics[width=65mm]{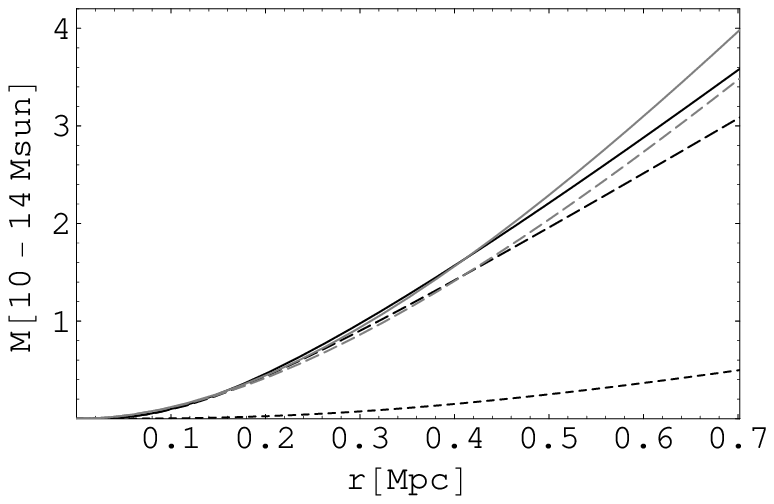}
\\
\\
\includegraphics[width=65mm]{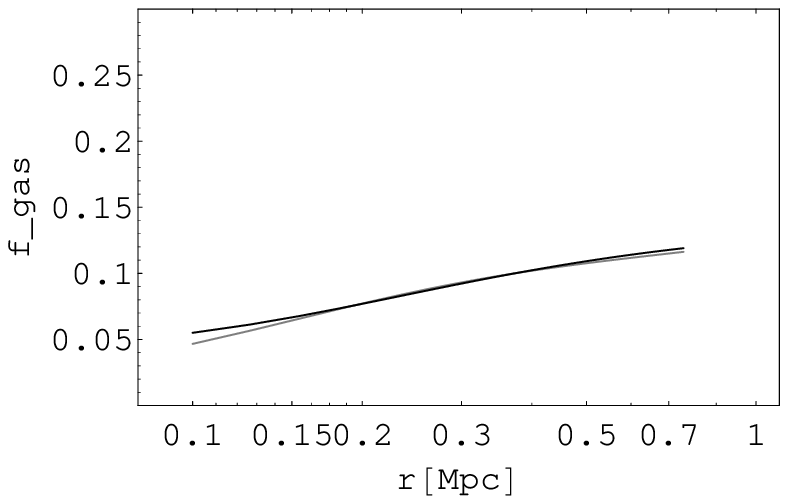} & \includegraphics[width=65mm]{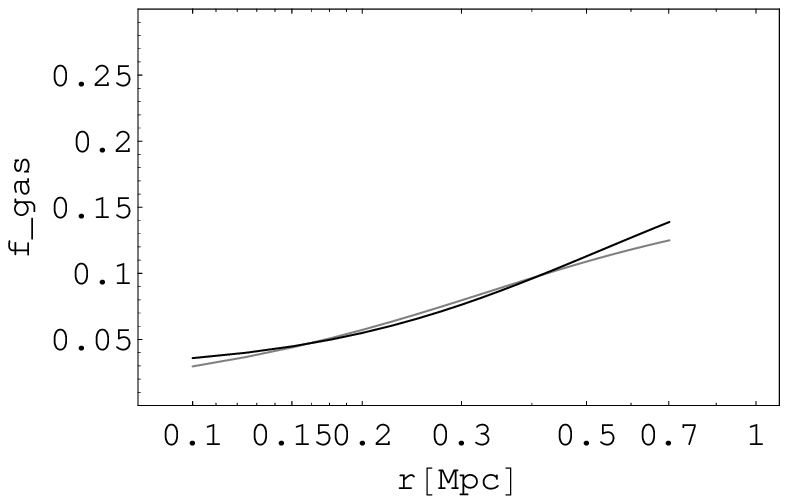}
\end{tabular}
\caption{(\emph{Top}) Mass profiles for the RIS model and the NFW
profile for A496 (\emph{left}) and Coma (\emph{right}). The
\emph{thick short-dashed line} is the mass $M_{X}(r)$ associated
with the ICM; the \emph{thick long-dashed-line} (RIS model) and
the \emph{thin long-dashed line} (NFW) are the mass $M_{DM} (r)$
associated with the dark matter; the \emph{thick solid line} (RIS
model) and the \emph{thin solid line} (NFW) are the total mass
profile $M_{tot} (r)$. On the vertical axis the mass is expressed
in units of $10^{14} \ {\rm{M}_{\odot}}$. (\emph{Bottom}) Gas
fraction $f_{gas}(r)$ for the RIS model (\emph{thick line}) and
the NFW profile (\emph{thin line}) for A496 (\emph{left}) and Coma
(\emph{right}). The figure is based on $h=0.5$.} \label{massprofiles}
\end{figure}

\begin{figure}[!t]
\begin{tabular}{c c}
\includegraphics[width=65mm]{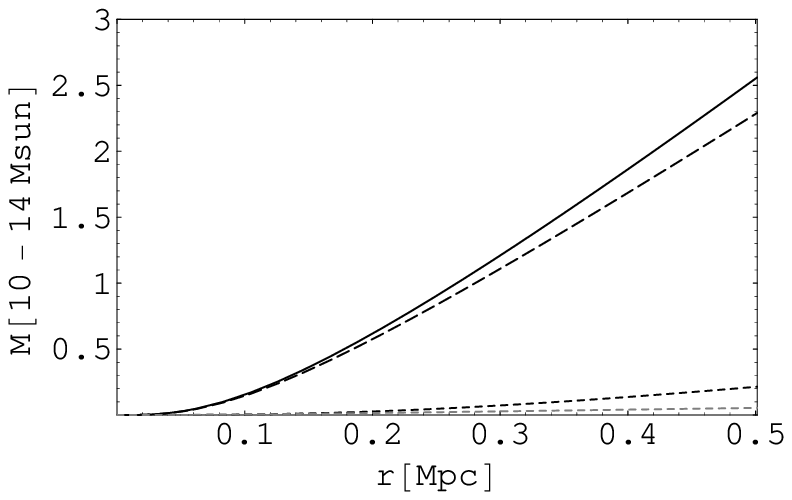} & \includegraphics[width=65mm]{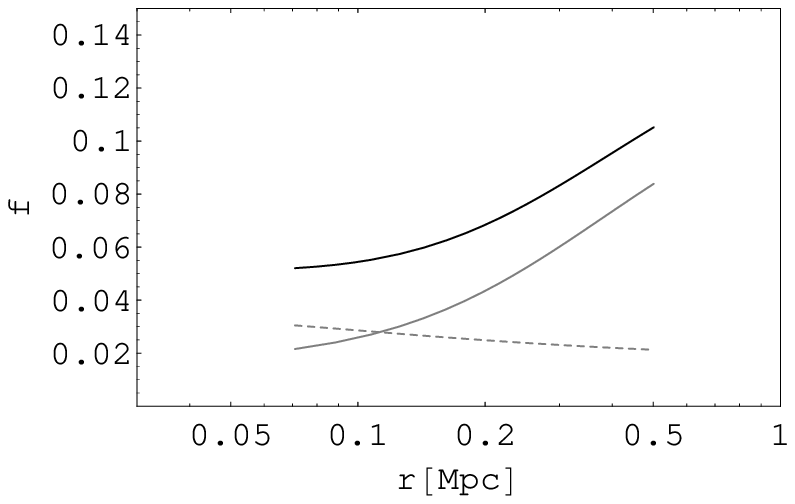}
\end{tabular}
\caption{(\emph{Left}) Mass profiles for the RIS model for Coma. The \emph{thin short-dashed line} is the mass $M_{G}(r)$ associated with the galaxies; the
\emph{thick short-dashed line} is the mass $M_{X}(r)$ associated
with the ICM; the \emph{thick long-dashed-line} is the actual mass $M_{DM}^{actual} (r)$
associated with the dark matter; the \emph{thick solid line} is the total mass
profile $M_{tot} (r)$. On the vertical axis the mass is expressed
in units of $10^{14} \ {\rm{M}_{\odot}}$. (\emph{Right}) Galaxy fraction $f_{G}(r)$ (\emph{thin dashed line}), gas
fraction $f_{gas}(r)$ (\emph{thin solid line}), and visible mass fraction $f_{G}(r)+f_{gas}(r)$ (\emph{thick solid line}) for the same RIS model for Coma. The figure is based on a Hubble constant corresponding to $h=0.7$.}
\label{Comaprofiles(h=0.7)}
\end{figure}

\end{document}